\documentclass[aps]{revtex4}

\begin{document}


%
%

\title{QUANTUM VERSION OF GAUGE INVARIANCE AND NUCLEON INTERNAL STRUCTURE}

\author{Fan Wang}

\address{Department of Physics, Nanjing University,
Nanjing, 210093, China}

\author{Wei-Min Sun}

\address{Department of Physics, Nanjing University, Nanjing,
210093, China}

\author{Xiao-Fu L\"{u}}

\address{Department of Physics, Sichuan University, Chengdu, 610064,
China}

\begin{abstract}
The conflict between canonical commutation relation and gauge
invariance, which both the momentum and angular momentum of quark
and gluon should satisfy, is clarified. The quantum version of
gauge invariance is studied.  The gauge independence of the matrix
elements of quark momentum and angular momentum operators between
physical states are proved. We suggest to use the canonical quark
momentum and angular momentum distributions to describe the
nucleon internal structure in order to establish an internal
consistent description of hadron spectroscopy and hadron
structure. The same problem for the atomic spectroscopy and
structure is discussed.

\end{abstract}

\maketitle

\section{Conflict between Canonical Commutation Relation and Gauge Invariance}    

The nucleon (atom) is a QCD (QED) gauge field system. The momentum
and angular momentum of the nucleon (atom) is the sum of
contributions from quark (electron) and gluon (photon)
respectively:
\begin{eqnarray}
\vec{P}&=&{\int}d^3x{\psi}^{\dag}\frac{\vec{\nabla}}{i}{\psi}+{\int}d^3xE_i{\vec{\nabla}}A_i.
\label{pequ}
\end{eqnarray}
\begin{eqnarray}
\vec{J}&=&\frac{1}{2}{\int}d^3x{\psi}^{\dag}\vec{\Sigma}\psi+
{\int}d^3x{\psi}^{\dag}{\vec{r}{\times}\frac{\vec{\nabla}}{i}{\psi}}+
{\int}d^3x{\vec{E}{\times}\vec{A}}+
{\int}d^3x{E_i\vec{r}{\times}{\vec{\nabla}}A_i} \label{jequ}
\end{eqnarray}
In the above equations, $\psi$ is the quark (electron) field,
$\vec{E}$ is the color electric (ordinary electric) fields,
$\vec{A}$ is the vector potential. In QCD case a summation over
color indices is understood. The good side of the above
decomposition is that each term in Eq.(1,2) satisfies the
canonical commutation relation of the momentum and angular
momentum operator, so they are quark and gluon momentum, quark
spin and quark orbital angular momentum, gluon spin and gluon
orbital angular momentum, respectively. However they are not gauge
invariant individually except the quark (electron) spin term.

Alternatively one can derive a gauge invariant decomposition,
\begin{eqnarray}
\vec{P}={\int}d^3x(\psi^{\dag}\frac{\vec{D}}{i}\psi+\vec{E}\times\vec{B}),
\end{eqnarray}
\begin{eqnarray}
\vec{J}={\int}d^3x(\frac{1}{2}\psi^{\dag}\vec{\Sigma}\psi+
\psi^{\dag}\vec{r}\times\frac{\vec{D}}{i}\psi+\vec{r}\times(\vec{E}\times\vec{B})).
\end{eqnarray}
The good side of this decomposition is that each term is gauge
invariant. However they do not satisfy the canonical commutation
relation individually except the quark (electron) spin
term\cite{cw}.

In classical gauge field theory, only gauge invariant quantities
are physically meaningful. In the study of nucleon internal parton
momentum and angular momentum structure, also only the gauge
invariant operators related to quark and gluon momenta, spin and
orbital angular momenta are appreciated\cite{ji}. In hadron
spectroscopy, partial wave analysis and multi-pole radiation are
widely used where the gauge non-invariant canonical momentum and
orbital angular momentum must be used accordingly.

Canonical momentum and orbital angular momentum have been used in
describing atomic structure for almost a century already. Are the
atomic electron momentum and orbital angular momentum not
measurable ones? Can these operators be used to describe the
nucleon internal structure? In this report we show that the
canonical quark (electron) momentum and orbital angular momentum
have gauge independent matrix elements between physical states and
so is observable, which should be used to establish an internal
consistent description of hadron spectroscopy and hadron internal
structure.

\section{Quantum Version of Gauge Invariance}
F.Strocchi and A.S.Wightman studied the quantum version of gauge
invariance\cite{sw}.

A gauge (or a quantization scheme) in a quantum gauge field theory
is specified by

(a) field operators: \emph{A$_{\mu}$},the gauge potential;
\emph{j$_{\mu}$}, the gauge interaction current; $\psi$, the
fermion field and other fields of the gauge in a Hilbert space
\emph{H};

(b) a representation \emph{U} of the Poincar$\acute{e}$ group in
\emph{H};

(c) a sesquilinear form (Gupta scalar product)
$\langle\Phi,\Psi\rangle$ on \emph{H} with respect to which
\emph{U} is unitary, $\Phi$ and $\Psi$ are vectors in \emph{H};

(d) a distinguished subspace $\emph{H}' \subset \emph{H}$ such
that

   (i) The restriction of the sesquilinear form to $\emph{H}'$ is
   bounded and nonnegative
\begin{center}
$\langle{\Psi} ,{\Psi}\rangle\geq{0}$ ~~~for${\Psi}\in\emph{H}'$.
\end{center}

   (ii) The analogue of the Maxwell equation holds in the sense
   that
\begin{equation}
\langle \Phi,(\partial_{\mu}F^{\mu\nu}-j^{\nu})\Psi \rangle=0
\end{equation}
for all $\Phi,\Psi \in \emph{H}'$.

   (iii) $\emph{H}'$ has a subspace $\emph{H}''$ consisting of vectors
   $\Phi$ in $\emph{H}'$ of zero length $\langle \Phi,\Phi \rangle=0$.
   The physical Hilbert space is $\emph{H}_{phys}=\emph{H}'/\emph{H}''$.

There exists a unique vector $\Psi_0$, called the vacuum, which is
invariant under the translation subgroup of the Poincar$\acute{e}$
group. The vector $\Psi_0$ lies in $\emph{H}'$.

This is a generalization of the Gupta-Bleuler quantization scheme
of QED.

A generalized gauge transformation is an ordered pair consisting
of two gauges
\begin{center}
$<{\emph{A}_{1\mu},\emph{H}_1,<\cdot,\cdot>_1,\emph{H}'_1}>$ and
$<{\emph{A}_{2\mu},\emph{H}_2,<\cdot,\cdot>_2,\emph{H}'_2}>$
\end{center}
together with a bijection \emph{g} of $\emph(H)_{1phys}$ onto
$\emph(H)_{2phys}$
\begin{center}
$[\Psi_2]=g[\Psi_1],[\Phi_2]=g[\Phi_1]$
\end{center}
\begin{center}
$[\Psi_{20}]=g[\Psi_{10}]$
\end{center}

Note that in the quantum version there is no need of the form of
classical gauge invariance, such as $F_{1\mu\nu}=F_{2\mu\nu}$.
Only under some special gauge transformation, one has such a gauge
invariant form. Instead the gauge invariance of an operator is
classified into four categories, i.e., gauge independence, weak
gauge invariance, gauge invariance, and strict gauge invariance.

An operator \emph{O}, mapping \emph{H} into \emph{H}, is called
gauge independent if
\begin{equation}
\langle\Phi,\emph{O}\Psi\rangle=\langle\Phi+\chi_1,\emph{O}(\Psi+\chi_2)\rangle
\end{equation}
for all $\Phi, \Psi \in \emph{H}'$ and all $\chi_1, \chi_2 \in
\emph{H}''$. In other words, the matrix elements $\langle\Phi,
\emph{O}\Psi\rangle$ for $\Phi, \Psi \in\emph{H}'$ depend only on
the equivalence classes $[\Phi], [\Psi] \in \emph{H$_{phys}$}$.
Such a gauge independent operator is an observable. The stronger
restricted operators, weakly gauge invariant, gauge invariant and
strictly gauge invariant ones are all gauge independent ones,
obviously they are observable. The classically gauge invariant
ones belong to the strictly gauge invariant category. Quantum
gauge field theory includes more observable operators than the
classical one.

\section{Quark (Electron) Momentum and Orbital Angular Momentum Are Observable }

To prove the quark (electron) momentum and orbital angular
momentum are observable, one has to prove they are gauge
independent, i.e., one has to prove
\begin{eqnarray}
\langle(\Phi+\chi_1),\vec{P}_q(\Psi+\chi_2)\rangle=\langle\Phi,\vec{P}_q\Psi\rangle,\nonumber\\
\langle(\Phi+\chi_1),\vec{J}_q(\Psi+\chi_2)\rangle=\langle\Phi,\vec{J}_q\Psi\rangle,
\end{eqnarray}
or equivalently to prove
\begin{eqnarray}
\langle\chi_1,\vec{P}_q\chi_2\rangle=0,~~~\langle\Psi,\vec{P}_q\chi_2\rangle=0,~~~\langle\chi_1,\vec{P}_q\Psi\rangle=0,
\end{eqnarray}
and the same for $\vec{J}_q$, where $\Psi\in\emph{H}'$  and
$\chi_1, \chi_2\in\emph{H}''$. All states $\chi$ of $\emph{H}''$
can be expressed as $\sum_n
a_n(\partial_{\mu}\emph{A}^{\mu})^n|\Psi_{phys}\rangle$, so one
has to consider
\begin{equation}
_{out}\langle\Phi|{\int}d^3y\psi^{\dag}(y)\frac{\vec{\nabla}}{i}\psi(y)\partial_{\mu}A^{\mu}(x)|\Psi\rangle_{in},
\end{equation}
which in the interaction representation can be written as
\begin{equation}
\langle\Phi|\emph{T}(\int
d^3y\psi^{\dag}(y)\frac{\vec{\nabla}}{i}\psi(y)\partial_{\mu}A^{\mu}(x)\emph{S})|\Psi\rangle,
\end{equation}
where $\emph{T}$ is time-ordering operator, $\emph{S}$ is the
scattering operator. Expanding the scattering operator as usual,
one has
\begin{equation}
\emph{S}=\sum_n \frac{(-i)^n}{n!} \int dx_1{\cdots}dx_n\emph{T}
\mathcal{H}_I (x_1) \cdots \mathcal{H}_I (x_n),
\end{equation}
where $\mathcal{H}_I(x)=-j_{\mu}A^{\mu}(x)$. The physical states
contain only transverse gluons (photons), so the operator
$\partial_{\mu}A^{\mu}(x)$ must be contracted with one
$A^{\nu}(y)$ in the interaction term $\mathcal{H}_I(x)$. This will
give rise to a term
\begin{equation}
\partial_{\mu}^x\overbrace{{A^{\mu}(x)A^{\nu}(y)}}=\partial_{\mu}^x\{g^{\mu\nu}D(x-y)\}=\partial^{\nu}_xD(x-y)
=-\partial^{\nu}_yD(x-y),
\end{equation}
here the symbol
$\overbrace{A^{\mu}(x)A^{\nu}(y)}=g^{\mu\nu}D(x-y)$ means
contraction. Then using integration by parts one can move the
differential operator $\partial_{\nu}$ to act on $j_{\nu}$ in the
interaction term $\mathcal{H}_I(y)$ and use current conservation
$\partial_{\nu}j^{\nu}=0$ to prove $Eq.(10)=0$. The other terms of
Eq.(8) can be proved to be zero in the same way. Thus one has
proved the quark (electron) momentum and orbital angular momentum
operators are gauge independent and so are observable.

\section{Conclusion}

The conflict between canonical commutation relation and gauge
invariance of the quark (electron) momentum and angular momentum
operators in a nucleon (atom) can be remedied by using gauge
non-invariant canonical quark (electron) momentum and orbital
angular momentum operators because they are gauge independent
ones. This problem has been discussed by us since 1998\cite{chen}.
Here we verified the results there by an alternative argument.

\section*{Acknowledgments}

This work is supported by NSFC through grant 10375030, 10435080.

\end{document}